\documentstyle[prl,aps,multicol,epsfig,amsmath]{revtex}
\draft \tolerance=10000

\def\vec#1{{\rm\bf #1}}

\begin{document}
\draft \twocolumn[\hsize\textwidth\columnwidth\hsize\csname
@twocolumnfalse\endcsname

\title{Photonic band structure of highly deformable, self-assembling systems}
\author{P.A. Bermel and M. Warner} \address{Cavendish
Laboratory, University of Cambridge, Madingley Road, Cambridge CB3
0HE, U.K.} \date{\today} \maketitle
\begin{abstract} We calculate the photonic band structure at normal incidence
of highly deformable, self-assembling systems - cholesteric
elastomers  subjected to external stress. Cholesterics display 
brilliant reflection and lasing owing to gaps in their photonic 
band structure. The band structure of cholesteric elastomers 
varies sensitively with strain, showing new gaps opening up and 
shifting in frequency. A novel prediction of a total band gap is 
made, and is expected to occur in the vicinity of the previously 
observed de Vries bandgap, which is only for one polarisation. 
\end{abstract}
\vspace{0.2cm} \noindent {PACS numbers:} 61.30.-v, 81.40.Jj,
42.70.Qs] \narrowtext

Photonic band-gap (PBG) materials offer a new approach to the manipulation
of light that depends on the structure rather than the atomic or
molecular properties of materials.  
These materials have two unique properties which has spurred 
interest in their design, namely the localization of light \cite{John:87}
and modification of the spontaneous emission spectrum from atoms and
molecules \cite{Tocci:96}.
Several approaches have been taken to manufacture PBG materials.
Yablonovitch constructed an fcc photonic crystal by drilling holes
into a dielectric medium \cite{Yablonovitch:89}.  Later, Ozbay
and co-workers designed a picket fence structure which is assembled
by stacking two-dimensional layers \cite{Ozbay:94}.

Recently, there has been an increased interest in self-assembling PBG
systems due to their relative ease of manufacture for 
operation at optical and near-infrared wavelengths. 
Several examples include air holes in a titania matrix 
\cite{Wijnhoven:98}, copolymer-homopolymer films which form 
lamellar structures \cite{Urbas:99}, thin films of PMMA infilled 
with $\mbox{SnS}_2$ \cite{Muller:00}, and cholesteric liquid crystals (CLC's)
\cite{deVries:51,deGennes:93,liquids,Palffy-Muhoray:01}.

One of the most promising applications of photonic band-gap materials
is in low-threshold lasing.
Yablonovitch \cite{Yablonovitch:89} first predicted that the lasing
threshold would be decreased by introducing a defect 
into an otherwise perfect photonic material.
Since spontaneous emission is suppressed in the bulk,
excitation would not be drained by any emission into non-lasing 
modes.  Such low-threshold lasing has recently been 
observed in two-dimensional photonic crystals \cite{Painter:99}.
Alternatively, one can design lasers that take advantage of the enhanced dwell 
time associated with the band edge divergence of the density of states
\cite{Dowling:94}.
Experimentally, this band-edge lasing has been observed in CLC's
\cite{liquids} and cholesteric elastomers (CE's) \cite{Palffy-Muhoray:01}.

A CLC has local orientational ordering
along a director $\vec{n}$, which rotates as a periodic
function of distance along the pitch axis $z$.  The director
of an ideal CLC advances uniformly, tracing out
a helix of pitch $p_0$.  The pitch can be adjusted
to match the wavelength of visible light, whereupon a number
of spectacular optical effects are observed experimentally
and explained theoretically \cite{deVries:51,deGennes:93}.  In particular,
in experiments conducted at normal incidence,
circularly polarized light which twists in the same sense as
the helix is reflected with its original polarization, while circularly
polarized light that twists in the opposite sense is transmitted unchanged.
Normal incidence has been of prime concern since the optical 
response of such twisting nematic media is the basis of liquid 
crystal (LC) display technology.
A CLC can be considered locally uniaxial, with a dielectric
permittivity $\epsilon_{\parallel}$ along $\vec{n}$ and 
$\epsilon_{\perp}$ perpendicular to $\vec{n}$.  By solving
Maxwell's equations in a rotating frame, de Vries found a single
band gap in the photonic structure of an ideal CLC at normal
incidence \cite{deVries:51}.  

The calculations we present on CE's point to new phenomena and new 
applications, not possible in existing photonics and hitherto 
unsuspected in the liquid crystal field.
For instance, we find multiple gaps, some not 
at the zone edges, in contrast to classical CLC's.  We also observe 
gaps for light of the opposite handedness to the underlying helix, 
again unexpected in classical CLC systems.  At some points the gaps 
for both polarizations overlap, giving a total gap of significance 
when polarisation control is required.  Our systems are highly 
deformable (to many 100s\%) and we shall find shifts in the 
(developing) band structure that can be large.  Existent photonic 
media typically have piecewise variation of an isotropic 
refractive index in going between a matrix and its inclusions.  By 
contrast, CLC's have a continuous variation of the 
principal axes of birefringence.  
Polarisation effects are thus 
very subtle and become more so for oblique incidence, which we 
consider in greater detail elsewhere.  
Control of polarisation is at the heart of LC and optical devices;
we thus view this work as a first step toward 
new classes of photonic solids with deformable, tunable band 
structures.


CE's can be made by
crosslinking cholesteric polymer liquid crystals
\cite{Finkelmann_first}. Defect-free monodomain rubber strips
tens of centimeters long display spectacular optical effects,
viz. large changes in the frequencies of reflection and lasing
\cite{Palffy-Muhoray:01} in response to imposed mechanical strains
that couple to director orientation (fig.~\ref{stretch}). These
strips can be thick and are oriented not by surface anchoring as
in liquids, but by interaction between local directors and the
rubber matrix.

Elongations $\lambda \equiv \lambda_{xx}$, applied perpendicularly
to the pitch axis
(see fig.~\ref{stretch}) are predicted to coarsen the initially

\begin{figure}[b]
\protect
\begin{center} \includegraphics[width=0.35\textwidth]{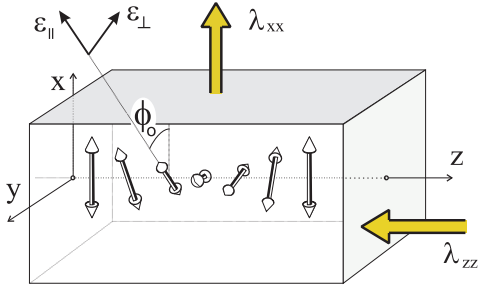}
\vspace{0.1cm} \caption{A CE showing the
initially helical director distribution, $\phi_0(z)$. Elongation
$\lambda_{xx}$ is applied perpendicular to the pitch axis which
contracts by $\lambda_{zz}$. The dielectric tensor is represented
in its local principal frame by $\epsilon_{\parallel}$ and
$\epsilon_{\perp}$. \label{stretch} }
\end{center} \end{figure}

\begin{figure}[t!]
\protect \vspace{0.1cm}
\begin{center}\includegraphics[width=0.4\textwidth]{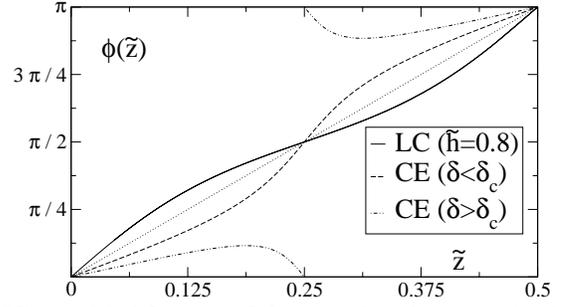}
\vspace{0.1cm} \caption{ Modification of director distribution,
$\phi(\tilde{z})$ by either mechanical strain $\delta$
(CE) or by DC magnetic fields $\tilde{h}$ (LC). The
dielectric
tensor's principal frame follows $\phi(\tilde{z})$;
 distortions to the helix then induce changes in the band structure.
\label{angles} }
\end{center}
\vspace{-0.6cm}
\end{figure}

\hspace{-0.34cm}helical director structure, given by $\phi_0 = q_{0} z$ (where
$\phi$ is the angle the director makes with the $x$-axis), to one
dominated by regions of slowly varying angles, separated by
increasingly sharp twist walls \cite{Warner:00}: see fig.~\ref{angles}.

At a critical $\lambda = \lambda_{c}$, the walls become
thermodynamically unstable and the director experiences periodic
oscillations about $\phi = 0$ which diminish with increasing
$\lambda$. There are attendant contractions perpendicular to the
stretch. The pitch shrinks affinely \cite{Palffy-Muhoray:01} with
the matrix. In the small stretching limit ($\lambda \rightarrow
1$), the pitch varies as $p = p_{0} \lambda^{-2/7}$ and the first
reciprocal lattice vector goes as $q = q_{0} \lambda^{2/7}$. Thus, 
the band structure changes upon extension because of two factors: 
the dilation of the reciprocal space and the change in the 
modulation character of the dielectric tensor along the pitch 
axis, $\underline{\underline{\epsilon}}( z )$.  These changes can 
be very large.

{\it Theory.}---Maxwell's equations yield \cite{Meade:93}:
\begin{equation}
\left( \frac{\omega}{c} \right) ^{2} \vec{H} = \vec{\nabla} \times
\left[ \underline{\underline{\epsilon}}(z)^{-1} \left(
\vec{\nabla} \times \vec{H} \right) \right] \; .\label{Max}
\end{equation}
Consider normal incidence, along the $z$ axis. The magnetic field
$\vec{H}$ is transverse and exists wholly in the $xy$ plane. We
thus suppress the $z$ components in the inverse dielectric tensor,
given by
\( \left(
\begin{array}{cc}\epsilon_{\parallel}^{-1} & 0\\ 0 &
\epsilon_{\perp}^{-1} \\
\end{array} 
\right) \)
in its principal frame (oriented at angle $\phi(z)$ to $x$,
 see fig.~\ref{stretch}), and $ \vec{\nabla}$ is only
$\left( d/dz \right) \hat{z}$.

We apply Bloch's theorem to decompose $\vec{H}$
into plane wave components \cite{Meade:93}, so that
\begin{equation}
\vec{H} = \sum_{G, \gamma} h_{(G \gamma)} \hat{e}_{\gamma} e^{ i
(k+G) z } \; ,\label{Bloch}
\end{equation}
where the unit vectors are \( \gamma = \left\{ 1, 2 \right\} \),
$\hat{e}_1 = \hat{x}$ and $\hat{e}_2 = \hat{y}$, and the
reciprocal lattice vector \( \vec{G} = 2 n q \hat{z} \), for $n$
integer. This procedure yields a matrix equation that reduces to a 
dimensionless form: lengths transform according to $z \rightarrow 
\tilde{z} = z q / 2 \pi $, wave vectors go as $k \rightarrow 
\tilde{k} = k / q$, reciprocal lattice vectors $G \rightarrow 2 
n$, frequencies go as
$\omega \rightarrow \tilde{\omega} = 
\omega/(c q \sqrt{a})$ and $\underline{\underline{\epsilon}}^{-1} 
\rightarrow \underline{\underline{\epsilon}}^{-1}/a $ where $a = 
\frac{1}{2}(\frac{1}{\epsilon_{\parallel}} + 
\frac{1}{\epsilon_{\perp}})$. This reduction is important for the 
proper interpretation of the shifts of band structure 

\begin{figure}[b!] \vspace{0.1cm} 
\begin{center}
\centerline{ \epsfxsize=0.4\textwidth \epsfbox{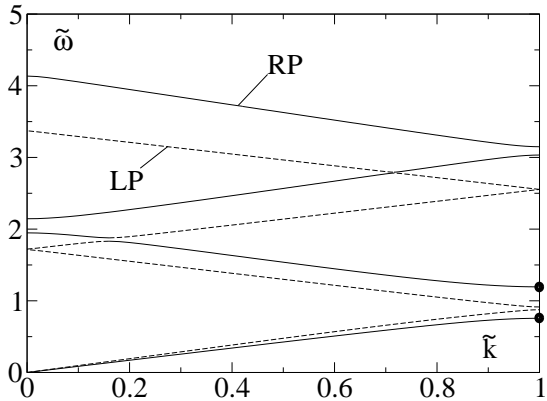}}
\vspace{0.1cm} \caption{ For \( \lambda = 1.1 < \lambda_c \), two
differences from the de Vries case  are observed: (1) additional 
bandgaps are created at higher zone boundaries for the RP branch 
and (2) bandgaps are also observed for the previously 
uninteresting LP branch, albeit much smaller than the RP gaps. The 
single gap in the de Vries case is approximately that marked by 
dots. \label{deVries:2} } 
\end{center}
\end{figure}

\begin{figure}[t!]
 \centerline{\epsfxsize=0.4\textwidth
\epsfbox{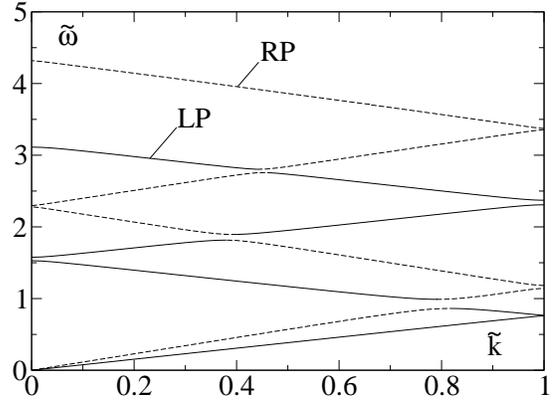}} \vspace{0.1cm} \caption{ For \(
\lambda =
1.3 > \lambda_c \), substantial divergence from the de Vries
dispersion relation is observed. A full
band gap away from the Brillouin zone boundary is observed,
as well as several anti-crossings between branches.
\label{deVries:3} }
\end{figure}

\hspace{-0.34cm}with 
elongation in figs. \ref{deVries:2} \& \ref{deVries:3}. Since $q = 
q_o\lambda^{2/7}$ changes with $\lambda$, so do $\omega$ and $k$. 

Eq. (\ref{Max}) then assumes the form 
$\underline{\underline{A}}^{\tilde{k}}_{(n \gamma), (n \gamma)'} 
h_{(n \gamma)'} = \tilde{\omega}^2 h_{(n \gamma)} $.
$\underline{\underline{A}}^{\tilde{k}}$ thus determines the
photonic
band structure of a CLC. It depends on the reduced inverse
dielectric tensor at arbitrary $z$ and thus angle $\phi = \phi
\left( z \right)$:
\begin{equation}
\underline{\underline{\epsilon}}^{-1} (z) =
\underline{\underline{1}} - \alpha \left[ (\cos 2 \phi)
\underline{\underline{\sigma}}_z + (\sin 2 \phi)
\underline{\underline{\sigma}}_x \right] \; , \label{epsilon_inv}
\end{equation}
where \( \alpha \equiv ( \epsilon_{\parallel} - \epsilon_{\perp} )
/ ( \epsilon_{\parallel} + \epsilon_{\perp} ) \) follows the
notation of de Vries \cite{deVries:51}, and the
$\underline{\underline{\sigma}}_i$ are the Pauli spin matrices.

One can then show that in $\gamma$ space,
$\underline{\underline{A}}^{\tilde{k}}$ is given by
\begin{eqnarray}
\underline{\underline{A}}_{n, n'}^{\tilde{k}} = (\tilde{k} &+& n
)^2 \delta_{n, n'} \underline{\underline{1}} \\ &+&\alpha (
\tilde{k} + n ) ( \tilde{k} + n' ) \left( c_{n'-n} \;
\underline{\underline{\sigma}}_z + s_{n'-n} \;
\underline{\underline{\sigma}}_x \right) \nonumber
\end{eqnarray}
where $s_{n'-n}$ and $c_{n'-n}$ are the Fourier coefficients of
$\sin (2 \phi)$ and $\cos (2 \phi)$, respectively: $s_n \equiv
\int^{1/2}_0d\tilde{z} \sin[2\phi(\tilde{z})]\exp(-4\pi i n
\tilde{z})$, and $c_n$ similarly.

The undeformed director angles $\phi (\tilde{z})$ are $\phi_0 = 2
\pi \tilde{z}$. On an $x$-strain $\lambda$, accompanied by
relaxation $\lambda_{yy} (\lambda)$ assumed to be uniform and
determined
by energy-minimisation, the principal frame orientation is given
by \cite{Warner:00}
\begin{equation}
\tan 2 \phi = \frac{2 \lambda \lambda_{yy} (r-1) \sin 4\pi {\tilde
z}}{(r-1)(\lambda^2+\lambda_{yy}^2) \cos 4 \pi{\tilde z} +
(r+1)(\lambda^2 - \lambda_{yy}^2)} \nonumber \; ,
\end{equation}
where $r$ is the shape anisotropy of the polymers underlying the
nematic phase. See fig.~\ref{angles} for $\phi (\tilde{z})$ for
various $\delta = \lambda -1$. From $\tan 2 \phi$, one can easily
obtain $\sin 2 \phi$ and $\cos 2 \phi$, and thus, $s_{n'-n}$ and
$c_{n'-n}$.

Numerical diagonalisation of the matrix
\(\underline{\underline{A}}_{(n \lambda), (n
\lambda)'}^{\tilde{k}}\) at a range of $\tilde{k}$ yields a
dispersion relation $\tilde{\omega}( \tilde{k})$, along with
eigenvectors giving the character of each solution. In general,
the eigenvectors are elliptically polarised inside the CLC
medium, with semi-major and semi-minor axes corresponding at each
point
to the local principal axes of the dielectric tensor,
and nearly circularly polarised {\it in vacuo} \cite{deVries:51}.
We take $\epsilon_{\parallel}=3$ and a depressed value
$\epsilon_{\perp}=1.2$ throughout, simply for readability.

At small $\tilde{k} \ll 1$, the dispersion relation for an ideal 
CLC (i.e., \( \lambda = \lambda_{yy} = 1 \)) is linear, 
corresponding to non-dispersive waves, with a simple effective 
refractive index \( m = \sqrt{(\epsilon_{\perp} + 
\epsilon_{\parallel})/2} \), suggesting that both modes 
effectively experience the same, homogeneous medium at long 
wavelengths. This small $\tilde{k}$ behaviour is initially 
retained in the strain-modified band structures, see 
fig.~\ref{deVries:2}. 

At $\tilde{k}=1$, the branch whose polarisation rotates in the 
same sense as the helix {\it in vacuo} develops the de Vries gap 
\cite{deVries:51}. The eigenmodes of this branch at the zone 
boundary are linearly polarised inside the CLC medium. The 
lower band's electric vector points along \( \vec{n} (\vec{r}) \), 
the upper band's perpendicularly to \( \vec{n} (\vec{r}) \), in 
the $xy$ plane. The other branch, however, cannot split 
analogously, since its polarisation {\it in vacuo} rotates in a 
sense opposite to the helix. This is qualitatively like the major 
gap in the distorted band structure, marked with dots at  
$\tilde{k} = 1$ in fig.~\ref{deVries:2}. Furthermore, no gaps are 
observed for $\tilde{k} > 1$, because band gaps are created only when 
degenerate energy states are linked by non-zero matrix elements. 
Since in the de Vries case, there are only two harmonic components 
of equal and 
opposite frequencies, only the matrix elements linking 
the two lowest energy states on either side of the first Brillouin 
zone boundary are non-vanishing. 

We now stretch CE's with $r=1.9$ for definiteness, which 
gives a \( \lambda_c \approx r^{2/7} \approx 1.2 \) 
\cite{Warner:00}. Fig.~\ref{deVries:2} shows the dispersion 
relation for an elongation \( \lambda = 1.1 < \lambda_c \). Since 
the first Brillouin zone boundary is at \( \tilde{k} = 1 \), band 
gaps may occur at \( k = n q_{0} \lambda^{\beta} \), for integer 
  $n \neq 0$, with $\beta = 2/7$. This corresponds to a shift in colour and
lasing frequency \cite{Palffy-Muhoray:01} toward the ultraviolet.

\vspace{-2cm}

\vspace{-0.25cm}

For $\lambda \geq \lambda_c$, there is a qualitative change in the
behaviour of the director, $\phi(z)$ (see fig.~\ref{angles}), and
thus a qualitative change in the band structure. Additionally,
the scaling behaviour of \( \lambda_{zz} (\lambda) =
\lambda^{-\beta} \) changes \cite{Warner:00} from \( \beta = 2 /
7 \), in the limit of small stretching ($\lambda \sim 1$), to \(
\beta = 1/2 \) for \( \lambda
> \lambda_c \), the classical exponent for
isotropic CE's. Fig.~\ref{deVries:3} shows the dispersion
relation for a stretch \( \lambda = 1.3 > \lambda_c \).

\begin{figure}[t!]
\vspace{-1cm}
\begin{center}
\centerline{ \epsfxsize=0.4\textwidth
\epsfbox{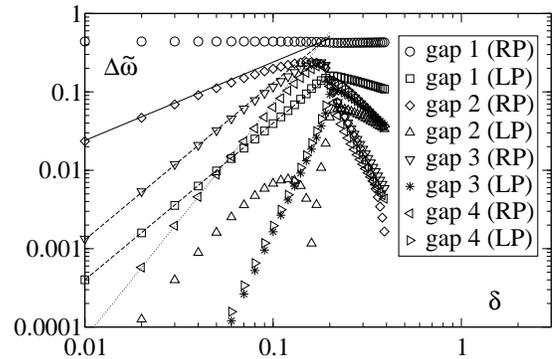}} \vspace{0.1cm}
 \caption{
Scaling of the first to fourth gap sizes of a CE
with \( \delta \). The points represent numerical
data, the straight lines, predictions from perturbation theory
(assuming a scaling form for $c_n$ and $s_n$). \label{gapscaling}}
\end{center}
\vspace{-0.8cm}
\end{figure}

We now analyse the gap structures that open up in the stretched
case, $\lambda > 1$. The elastic strain, $\delta \equiv \lambda -
1$, is the perturbation parameter modifying the perfect helical
structure. Whereas before, \( c_{\pm 1} = 1/2 \), and \( c_n = 0
\) otherwise, we now have non-zero values for \( c_{\pm n} \)
which scale as $\delta^{n-1}$ for $ n > 1 $ and $\delta \ll 1$; \(
c_{\pm 1} = 1 - {\mathcal O} \left( \delta^2 \right) \) and \(c_0
\sim \delta\). Applying degenerate perturbation theory, we
eliminate all matrix elements in \(
\underline{\underline{A}}^{n_0}_{(n \lambda), (n \lambda)'} \)
except for those linking the degenerate energy states, and predict
that the gaps for the interesting polarisation in the de Vries
case will scale with the magnitude of the off-diagonal elements,
given by $c_{n_0}$. The $1^{\mbox{st}}$ order (de Vries) gap will
be approximately constant, the $2^{\mbox{nd}}$ order gap (first
new gap) will grow as $\delta$, the $3^{\mbox{rd}}$ order gap will
grow as $\delta^2$, and so on: see fig.~\ref{gapscaling}. More
precisely, at the $1^{\mbox{st}}$ order gap, \( \tilde{\omega}^2 =
1 \pm \alpha \left( c_1 \pm \sqrt{c_0^2-s_1^2} \right) \), since
$s_0 = 0$.
As a result, we also
predict that the opposite polarisation will now have a non-zero
gap that scales as $\delta^2$. That tells us that a {\it full}
photonic band gap will be created at $\tilde{k}=1$ as we perturb
our helix by transverse elongation, and that its size will scale
as $\delta^2$, the size of the much smaller gap of opposite
polarisation. See figs. \ref{deVries:2} and \ref{gapscaling}.

Of experimental interest is the {\it in vacuo} wavelength, 
$\Lambda$, of the light corresponding to a given $\tilde{\omega}$ 
on the dispersion relation, particularly at the gaps.
The definitions below (\ref{Bloch}) give $\Lambda 
= p_0/(\tilde{\omega} \sqrt{a} \lambda^{2/7})$.
Pitches $p_0$ typically 
give a band in the visible so the initial wavelengths are 
$\Lambda_0 = p_0 /\sqrt{a} \sim 500 {\rm nm}$ at 
$\tilde{\omega} = \lambda = 1$, which allows us to write
$\Lambda = \Lambda_0/(\tilde{\omega}\lambda^{2/7})$.  Likewise the first 
order de Vries gap
is given by $\Delta \Lambda \approx \Lambda_0/\lambda^{2/7} \alpha$.
The higher order gaps of
the same polarization will have widths of $\Delta \Lambda_n \approx
C \Lambda_0 \delta^{n-1}/(n \lambda^{2/7})$, where $C$
is a pre-factor of order unity that will depend on $r$ and $\alpha$.
For example, the second order gap in a rubber with $r=1.9$,  
$\alpha=0.43$, and
$\delta=0.1$, the gap will be $\Delta \Lambda_2 \approx 0.045 \Lambda_0$.  For
$\Lambda_0 \approx 800 {\rm nm}$, that implies a stop band for the
light with a circular polarization of the same sense as the helix will be
observed for $\Lambda = 362 \; {\rm nm}$ to $\Lambda = 398 \; {\rm nm}$.

Finally, a note about oblique incidence in CE's.  By
symmetry, we expect that the magnetic field must have
the same magnitude at all points for a given $z$, and only differ by
a phase.  That lets us generalize the $H$ vector for normal incidence, 
eq. (\ref{Bloch}), by $\vec{k} \rightarrow \vec{k}'=k_{\perp} \hat{\rho} + 
k_{\parallel} \hat{z}$ and 
$\hat{e}_{\gamma} \rightarrow \hat{e}_{(G\gamma)}'$.
Our preliminary findings indicate that for a constant $k'$, 
the stop bands shift
upwards as we increase the angle of incidence from zero, which implies
that refraction out of a normally incident beam
path is forbidden for modes just above the stop band.  That 
observation provides a mechanism to explain the spatial coherence of the light
produced by dye-doped pumped lasers based on CLC's
and CE's \cite{liquids,Palffy-Muhoray:01}.

\noindent{\it Cholesteric Liquids.}---We apply an external 
magnetic field $H$ along the $y$ direction. A CLC has 
an anisotropic susceptibility 
$\chi_a=\chi_{\parallel}-\chi_{\perp}$. The helix untwists 
(increasing the period) and coarsens \cite{deGennes:68} until the 
energy gain from aligning with the field balances the Frank 
penalty for deviations from the original structure. The coarsening 
of the director orientation is illustrated in fig.~\ref{angles}, 
with the $z$-coordinate being reduced to $\tilde{z}$ by the 
lengthening period. At a critical field $H_c = (\pi q_0/2) 
\sqrt{K_{22}/\chi_a}$, where $K_{22}$ is the Frank twist 
elastic constant, the period diverges logarithmically as the 
entire sample aligns with the external field \cite{deGennes:68}. 
For typical cholesteric liquids with a pitch of $20 \; {\rm \mu m}$, 
$H_c = 15,000\;{\rm G}$ and $E_c = 50\;{\rm sV/cm}$ \cite{deGennes:93}. 
Differences from the case of a CE
under strain are detailed in 
\cite{Warner:00}.

The optical implications of coarsening were investigated by
Meyer \cite{RBMbands}. The Fourier coefficients
describing the twist of the cholesteric liquid are found to scale just like the
coefficients describing CE's for \( \lambda <
\lambda_c \). 
One can use the results for the CE case for a cholesteric
liquid under the transformation \( \delta \rightarrow \tilde{h}^2\). 
The dispersion relations are qualitatively the
same as in fig.~\ref{deVries:2}. The width and scaling of the gaps created 
resemble those of
fig.~\ref{gapscaling}, but with all effects ending abruptly at
$\tilde{h} = 1$.

{\it Conclusions.}---An entirely new type of photonic material has
been described and characterized. Not only is it self-assembling
and easily available as large, defect-free
single crystals, but it
is highly deformable. Earlier descriptions
\cite{Warner:00,deGennes:68} of its modified periodic dielectric
structure have been used as the basis for calculating its band
structure. New gaps arise and their widths scale in a
well-understood fashion with the stretch applied to the material
or the strength of the external field. The midgap
frequencies shift position by large amounts comparable

\hspace{-0.34cm}to their initial values. 

We thank A Genack and P Palffy-Muhoray for introducing us to
lasing in cholesterics, and acknowledge valuable discussions with
EM Terentjev, Y Mao, H Finkelmann, ST Kim, W Stille, S
Shiyanovskii, PD Haynes, PB Littlewood and S Johnson.
\vspace{-0.5cm}


\end{document}